\documentclass[%
 reprint,
superscriptaddress,
 amsmath,amssymb,
 aps, prl,
]{revtex4-2}

\usepackage{graphicx}
\usepackage{dcolumn}
\usepackage{bm}
\usepackage{algorithm}
\usepackage{algorithmicx}
\usepackage{algpseudocode} 
\usepackage{siunitx}
\usepackage{subfig}
\usepackage{lipsum}
\usepackage{comment} 
\usepackage{mathtools} 
\usepackage{array} 
\usepackage{booktabs}
\usepackage{colortbl}
\usepackage[table,xcdraw]{xcolor}

\usepackage{color}

\begin{document} 

\pagenumbering{arabic}

\title{Navigating Phase Transitions with Path-Finding Algorithms: A Strategic Approach to Replica Exchange Monte Carlo}
\author{Akie Kowaguchi}
\affiliation{Department of Mechanical Engineering, Keio University, Yokohama 223-8522, Japan}
\affiliation{Cygames Research, Cygames Inc., Shibuya, Tokyo, Japan}

\author{Katsuhiro Endo}
\affiliation{Research Center for Computational Design of Advanced Functional Materials, National Institute of Advanced Industrial Science and Technology (AIST),
1-1-1 Umezono, Tsukuba, Ibaraki, 305-8568, Japan}

\author{Kentaro Nomura}
\affiliation{Preferred Networks, Inc.
1-6-1 Otemachi, Chiyoda, Tokyo, Japan}

\author{Shuichi Kurabayashi}
\affiliation{Cygames Research, Cygames Inc., Shibuya, Tokyo, Japan}
\affiliation{Graduate School of Media and Governance, Keio University, Fujisawa, Kanagawa, Japan, }

\author{Paul E. Brumby}
\affiliation{Department of Mechanical Engineering, Keio University, Yokohama 223-8522, Japan}
\author{Kenji Yasuoka}
\affiliation{Department of Mechanical Engineering, Keio University, Yokohama 223-8522, Japan}

\begin{abstract}
The replica exchange method is a powerful tool for overcoming slow relaxation in molecular simulations, but its efficiency depends strongly on the choice of the number and interval of replicas and their exchange probabilities. Here, we propose a new optimization scheme based on the Dijkstra algorithm that constructs an optimal exchange path by representing replicas and their exchange probabilities as a graph. Inspired by path-finding techniques widely used in computer science, including applications in game algorithms, our approach ensures that transitions follow a minimum entropy gradient path and effectively speeds up sampling even in systems exhibiting slow relaxation near critical points or phase transition regions. The method provides a systematic way to improve replica exchange efficiency and offers new insights into the control of relaxation dynamics, as demonstrated through applications to the solid-liquid phase transition of the Lennard-Jones bulk system.
\end{abstract}

\maketitle

\begin{figure}[h]
\includegraphics[width=8.4cm]{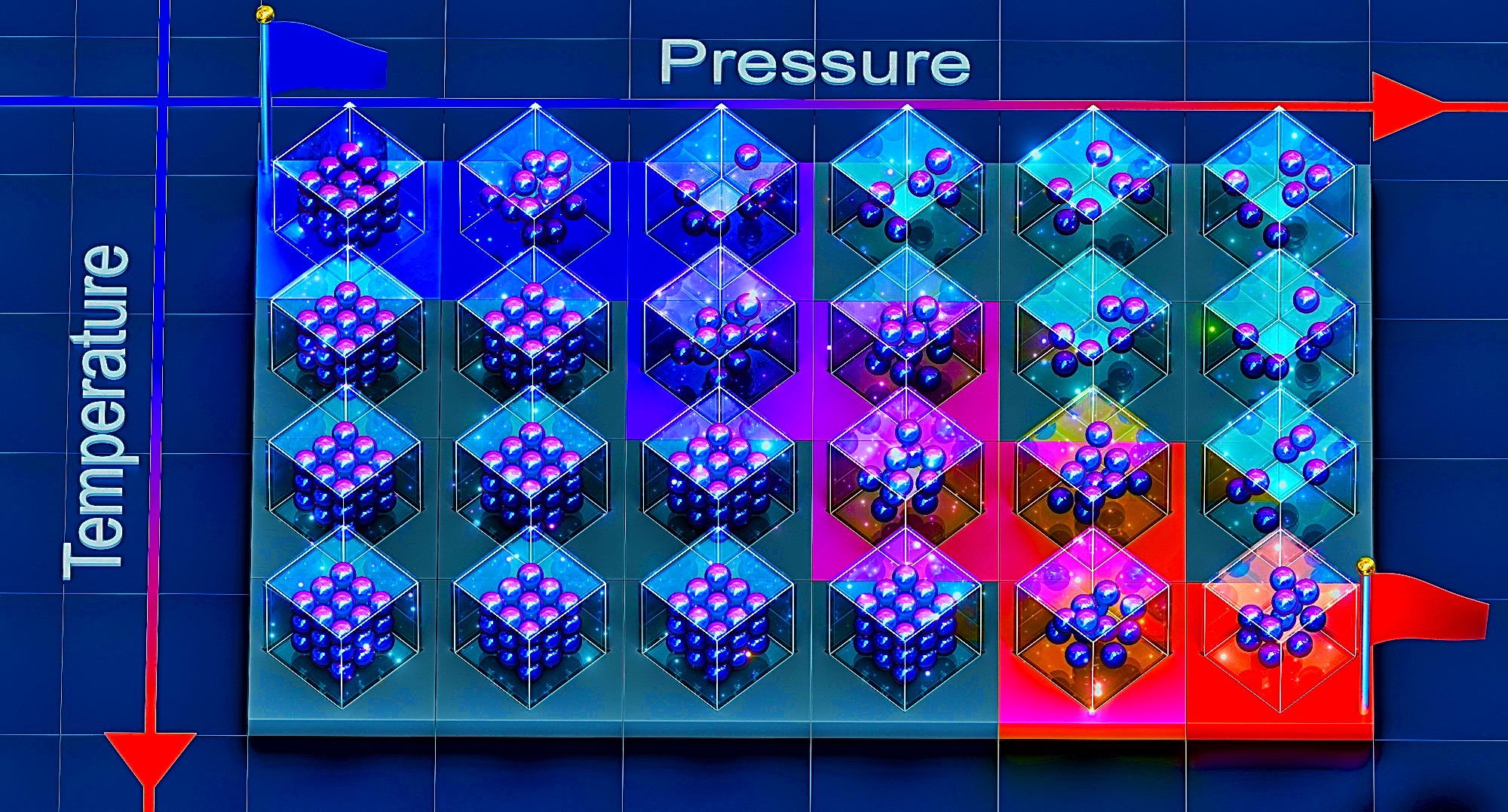}%
\label{toc}
\end{figure}

\section{Introduction}
In complex systems with high energy barriers, achieving the right balance between accurately capturing the true phase equilibrium states - critical transition states - and computational efficiency is necessary. Molecular dynamics (MD) simulations allow detailed tracking of the motion of atoms and molecules, but their application is restricted to short timescales. In contrast, Monte Carlo (MC) simulations enable the search for global energy minima using statistical random sampling while disregarding time-dependent dynamics. While MC simulations are a powerful tool for solving multidimensional problems, they face challenges in systems exhibiting phase transitions due to the long computational time required for the system to reach equilibrium. Therefore, to efficiently transition between these states of interest, a Replica Exchange Monte Carlo (REMC) method (Parallel Tempering, Metropolis-coupled Markov Chain Monte Carlo) \cite{hukushima1996exchange}, one of a number of extended ensemble methods \cite{berg1991multicanonical, berg1992multicanonical, lyubartsev1992new,marinari1992simulated, wang2001determining,mitsutake2010multi} (for reviews see, e.g., Refs.\cite{henin2022enhanced,earl2005parallel}), is employed due to its intuitive and straightforward approach. The method uses a Markov Chain Monte Carlo (MCMC) method to sample simultaneous probability distributions, which are a composite of multiple distributions with different parameter values. \par
This study aims to efficiently find the optimal transition pathways between any two states by selecting the appropriate replica parameters in the replica exchange method. It is also interesting to overcome the difficulty of replica exchange near phase transitions and to improve the technique's reproducibility and overall computational efficiency.
A specific example of its use is in predicting the first-order solid-liquid phase transition of the 12-6 Lennard--Jones system. The Lennard--Jones fluid is widely used as a fundamental force field in molecular simulation and is used as the basis of a large number of atomic pair-potentials. Still, studies on its phase diagram have been carried out even in recent years (for reviews see, e.g., Ref.\cite{schwerdtfeger2024100}), as it can reproduce universal solid-liquid behavior, despite its simple mathematical formulae. \par
The use of the replica exchange method is challenging near phase transitions where the specific heat diverges, as the exchange probability is determined by the distance between, and the width of, the two replica's probability distributions and their degree of overlap. If the parameters are not selected appropriately, localized areas with frequent exchanges and areas with little or no exchanges may occur, making it hard to draw out the benefits of this method. How to set the parameters used in the replica exchange method is still an area of active debate, with various considerations being made in terms of the frequency of exchanges, parameter intervals, and the number of replicas \cite{kofke2002acceptance, kofke2004comment, kofke2004erratum,rathore2005optimal,kone2005selection,schug2004all,katzgraber2006feedback,yan2000hyperparallel,sugita2000multidimensional,ogita2025atomistic}. \par
Equispaced replica arrangements were found to be inappropriate when applied to the strongly first-order nematic-solid phase transition region of some liquid crystal systems, which exhibit strongly first-order phase transitions in which the specific heat diverges in the $\beta=1/k_{B} T$ axis \cite{kowaguchi2021phase}. To address this issue, we developed a method that incorporates an evolutionary strategy for a fixed number of replicas and a parameter range such that the exchange probabilities all maintain a preset minimum value for any range \cite{kowaguchi2022optimal}. As a result, by applying an evolution strategy \cite{rechenberg1965cybernetic,schwefel1965kybernetische} to the Hamiltonian replica exchange method \cite{sugita1999replica}, exchange attempts with divergent specific heats in the $\beta$ axis could be avoided by tuning the replica spacing in the anisotropic term direction. Thus, the hysteresis phenomenon of the coarse-grained Hess-Su liquid crystal model \cite{hess1999pressure,steuer2003pressure,steuer2004phase} was eliminated, and accurate phase transition points were obtained, irrespective of the initial conditions used \cite{kowaguchi2024hysteresis}.\par
In this study, we extend this method, by removing the constraint that the number of replicas is predetermined and develop a process to find the number of replicas between which all exchanges can occur with a particular exchange probability or higher for an arbitrary parameter range. Traditionally, replica exchange in two dimensions is generally performed on a square grid, but there are alternative approaches to improve the exchange efficiency through distortion of the square grid near phase transitions, as in the work of Kimura et al. \cite{kimura2016extension}. Here, however, we implement the so-called king's grid, wherein the diagonal direction is added and we explore the question of whether a square grid is the more favorable exchange geometry for the replica exchange method. Next, after examining the impact of the grid configuration on the replica's random walk, the aim is to find the path on a graph where the replicas are treated as nodes and the exchange probability is considered as the distance (i.e., the weight of the edges), such that the number of edges is minimized. In addition to the motivation of finding the optimal spacing for the exchange axis in two dimensions, our goal is to identify a path that connects both the solid and liquid phases of the Lennard--Jones potential with high exchange probabilities, thereby enabling exploration between different phases with relatively fast relaxation times.\par

Somewhat surprisingly, there is great potential to apply methods and techniques from graph theory and computer science, which have been extensively used for solving path-finding problems in computer games, to find optimal routes across free energy landscapes in molecular simulations. There has been extensive research on search algorithms for game development in computer science. The simplest examples are the depth-first \cite{tarjan1972depth} and the breadth-first search methods \cite{bundy1984breadth}, which perform the entire search from the starting node to the goal. However, neither of these considers edge weights, which are useful when applied to maze problems. Thus, the Dijkstra method \cite{dijkstra2022note} was developed, which is more appropriate for the shortest path problem in weighted graphs. The aim is to find the path with the minimum total edge weight in a graph with fixed starting and ending points. By considering replicas as nodes and exchange probabilities as distances, i.e., weights of edges, replica selection becomes intuitive and paths can be mathematically determined in a rigorous manner. In a further development, the A* algorithm \cite{hart1968formal} introduces a heuristic function to preferentially search for the route closest to the goal while predicting possible routes. However, this latter method will not be used in the current study.\par

\section{Methods}

\subsection{Lennard--Jones Potential}  
In this work we select the commonly used Lennard--Jones potential to represent $256$ interacting particles:
\begin{equation}
U_{LJ} = 4\varepsilon\left[\left(\frac{\sigma_{ij}}{r_{ij}}\right)^{12} - \left(\frac{\sigma_{ij}}{r_{ij}}\right)^6\right],
\end{equation}
where $U_{LJ}$ is the potential energy between particle $i$ and $j$ at a distance $r_{ij}$, $\varepsilon$ is the depth of the potential energy wall, and $\sigma_{ij}$ is the distance between the two particles for which $U_{LJ} = 0$. \par

\subsection{Replica-Exchange Method}
This study employs the extension of the REMC to isobaric-isothermal ensembles proposed by Okabe et al. \cite{okabe2001replica}. 
We consider a simultaneous probability distribution of $M$ independent distributions subjected to different inverse temperatures $\beta_i(i=1,...,M)$ and pressures ${P_i}^* (i=1,...,M)$ given as
\begin{equation}
\Pi \lparen \lbrace X,\beta,P^*\rbrace \rparen=\prod_{m=1}^{M}\Pi_m \lparen X_m,\beta_m,P^*_m\rparen
\end{equation}
with alternating MCMC calculations for individual distributions and stochastic exchange attempts between neighbouring replicas.
This approach assumes that replica $m$ is in state $X$ and replica $n$ is in state $X'$. Attempts are made to exchange between neighboring systems in temperature and pressure, with the probability of acceptance for exchange between systems $m$ and $n$ determined as follows:
\begin{eqnarray}
\label{repequation}
W \left( X,\beta_m,P_{m}^{*} \mid X',\beta_n,P_{n}^{*} \right) = 
 \begin{cases}
 1 & \mathrm{for} \ \Delta \leq 0\\
 \exp(-\Delta) & \mathrm{for} \ \Delta >0 \ ,
 \end{cases}
\end{eqnarray}
where $\Delta$ is defined as
\begin{equation}
\Delta=(\beta_m-\beta_n)(U_{n}^{*}-U_{m}^{*})+(\beta_m P_{m}^{*}-\beta_n P_{n}^{*})(V_{n}^{*}-V_{m}^{*}) \ .
\end{equation}

For the $(m,n)$ replica, the symbols $\beta_{(m,n)}$, $V_{(m,n)}^{*}$, $P_{(m,n)}^{*}$, and $U_{(m,n)}^{*}$ denote the inverse temperature, volume, pressure, and internal energy, respectively.

\subsection{Graph Representation and Path Optimization}

As discussed in the introduction, search algorithms can provide us with a range of useful tools with which to overcome the difficulties presented by phase transitions. The Dijkstra algorithm \cite{dijkstra2022note} is a path-finding algorithm used to find the shortest path between specified nodes in a weighted graph. The algorithm starts by initially assigning an infinite distance to all nodes except the starting node. It then selects the node with the shortest known distance, explores adjacent nodes, and updates their distances if shorter paths through the current node are found. This process continues until the shortest path from the starting node to every node in the graph is determined (Supplementary S1). \par
\begin{figure}[t]
\includegraphics[width=8.4cm]{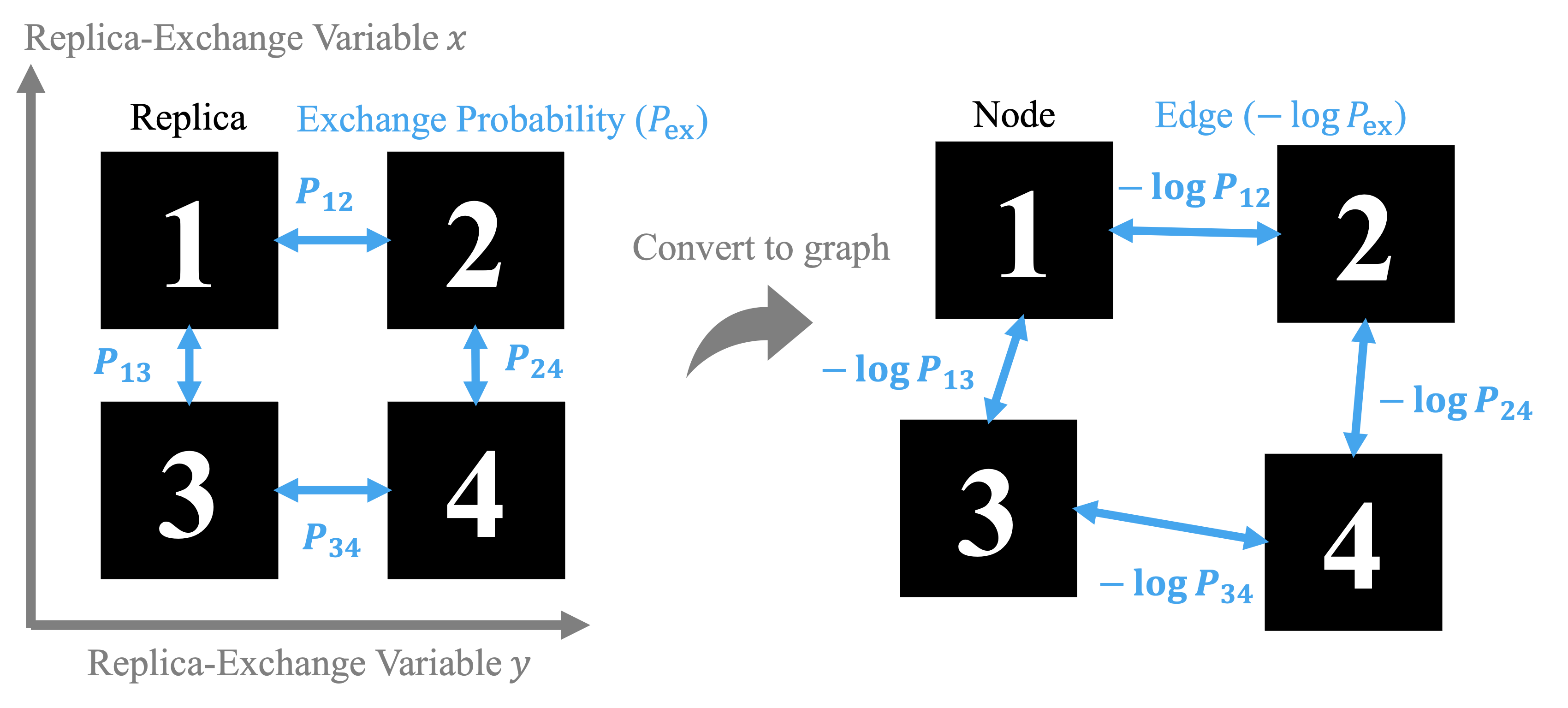}
\caption{(left) Edge and node definition procedure based on the example of a $2 \times 2$ arrangement with four replicas. The vertical axis represents the variable $x$ exchanged by the replica exchange method, and the horizontal axis represents the variable $y$. (right) Conversion to a graph with the exchange probabilities $P_{\rm{ex}}$ as distances $-\log P_{\rm{ex}}$.}
\label{dijkstra1}
\end{figure}

This study defines an undirected graph with each replica as a node and the exchange probability between replicas as an edge value, as shown in Fig. \ref{dijkstra1} (right). First, exchange probabilities are computed over a fine grid of temperature and pressure conditions to obtain a comprehensive probability landscape. These probabilities are then converted into distances using their logarithms, leveraging the additive properties of logarithms to represent the multiplicative nature of exchange probabilities, as shown in Fig. \ref{dijkstra1} (left).  
To construct the optimal path, edges are selectively retained based on the exchange probability: edges with probabilities lower than $10 \%$ are disconnected, while edges are iteratively extended until an exchange probability of at least $30 \%$ is reached. Additionally, weightings are applied to each edge to prioritize paths with fewer exchanges when multiple options with similar exchange probabilities exist. This is achieved by assigning a weight of $-\log{P_{ex}}+\alpha$, where $\alpha$ acts as a penalty to prefer paths with fewer steps. Here, $\alpha$ is set as a minimal constant. The adjacency matrix represents all edges connecting nodes in four directions, both horizontally and vertically.

\subsubsection{Physical Interpretation of the Optimal Path}

Kofke analysed the average exchange probability $\bar{P}_{acc}$ and argued that this quantity is related to the difference in entropy of the two systems in the canonical ensemble \cite{kofke2002acceptance,kofke2004comment,kofke2004erratum}.
The average exchange probability $\bar{P}_{acc}$ can be described as
\begin{equation}
\begin{aligned}
\bar{P}_{acc} \propto & \frac{\exp(-\Delta S/k)}{(\pi C)^{0.5}} 
\left[ \frac{4}{(1+B)^2} \right]^{C+1} \\
& \times \frac{1+B}{1-B}(1+O(C^{-0.5})), \quad C\rightarrow \infty.
\end{aligned}
\label{entropy}
\end{equation}
Here, $\Delta S$ is the entropy difference which can be expressed as $\Delta S/k=-C_V/k\ln(\beta_1/\beta_0)$ with the temperature ratio $B \equiv\beta_1/\beta_0<1$ and $C$ the constant-volume heat-capacity $C \equiv C_V/k$ in units of the Boltzmann constant $k$. Thus, a uniform exchange probability indicates that the entropy gradient is properly adjusted with respect to temperature and pressure. The optimal path will avoid abrupt free energy changes at phase transitions and allow for overall stable exchanges.\par

\subsubsection{Entropy Calculation Using WHAM}

To quantitatively evaluate the entropy landscape along the optimal path, we compute the entropy using the Weighted Histogram Analysis Method (WHAM). This method allows us to estimate the density of states $n(U^*, V^*)$, following the approach developed by Okumura et al. \cite{okumura2004molecular,okumura2006multibaric}. Given that entropy is a logarithmic function of the number of states, we calculate it as
\begin{equation}
S^*(U^*,V^*)=\ln n(U^*,V^*).
\end{equation}
By applying WHAM, we obtain a detailed entropy profile along the optimized exchange path.

\section{Results}
\subsection{Exploring Replica Exchange on Conventional Grids}

\begin{figure}[h]
\includegraphics[width=9.4cm]{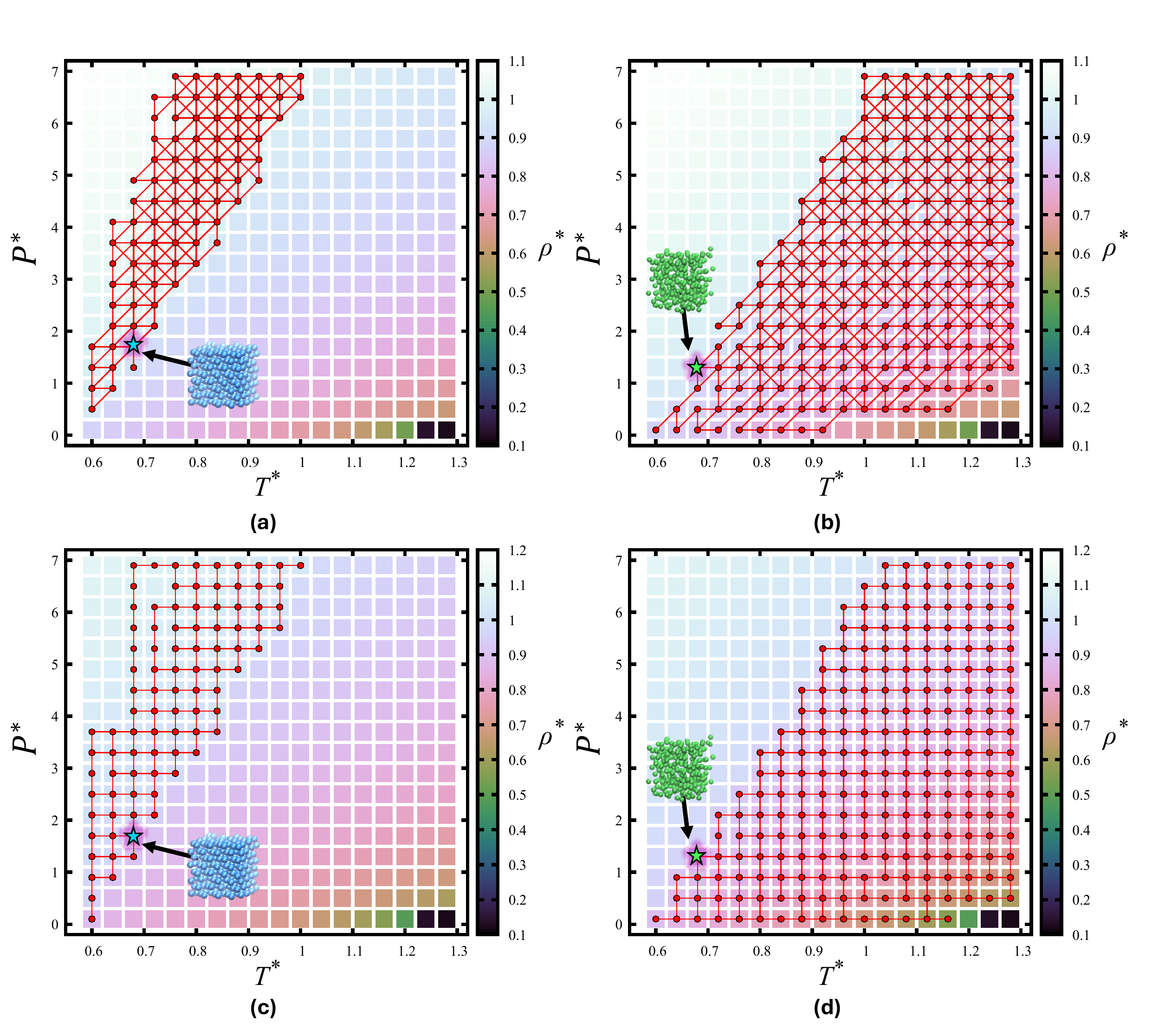}%
\caption{Trajectory of a single replica near the solid-liquid phase transition, superimposed on a density heatmap. 
The heatmap represents density $\rho^*$ as a function of temperature $T^*$ and pressure $P^*$, with the transition from high to low density (bottom-left to top-right) corresponding to the solid-liquid phase boundary. 
(a) and (b) show trajectories on the King’s grid, while (c) and (d) show trajectories on the Square grid. 
(a) and (c) correspond to a solid-phase state at $P^* = 1.7$, $T^* = 0.68$, while (b) and (d) correspond to a liquid-phase state at $P^* = 1.3$, $T^* = 0.68$. 
The results illustrate how replicas exhibit constrained random walks, reflecting the difficulty of crossing the phase boundary using conventional replica exchange.}
\label{result1}
\end{figure}

Before implementing Dijkstra's algorithm, we compare the impact of different grid configurations, specifically the Square grid and King's grid, on the trajectory of the replica's random walk. (The computational conditions are shown in Supplementary Table S1.) For a range of temperatures and pressures, examples of replica trajectories on an $18 \times 18$ grid, with variations in initial conditions and grid configurations, are presented in Fig. \ref{result1}. In Fig. \ref{result1}, we track the trajectory of single replicas who's initial positions are located near to, and on either side of, the phase transition line. Their initial conditions are $P^* = 1.7$, $T^* = 0.68$ for the solid phase replicas, and $P^* = 1.3$, $T^* = 0.68$ for the liquid phase replicas. The trajectories for the solid phase replicas are shown in (a) and (b), and those for the liquid phase replicas are shown in (c) and (d). Furthermore, the trajectories for the Square grid are shown in (a) and (c), and those for the King's grid are shown in (b) and (d). From (a) and (b), it is evident that in the solid phase, the random walk is confined to the high-density ($\rho^*$) region and cannot sample across the diagonal phase boundary line. Likewise, from (c) and (d), it can be seen that in the liquid phase, the random walk is confined to the low-density region and also cannot sample across the diagonal phase boundary line. (For a comprehensive analysis of replica trajectories across all initial phases, see the Supplementary S2.)
Based on these trajectory results, the superiority of the King's grid with diagonal exchanges cannot be demonstrated. Therefore, if we wish to sample different solid and liquid phases using a single replica exchange method, it is necessary to extract the optimal parameter spacing that ensures sufficiently high exchange probabilities and determine an appropriate parameter spacing for performing random walks between the two phases.

\subsection{Efficient Exchange Pathways Across Solid, Liquid, and Gas Phases}

\begin{figure}[h]
\includegraphics[width=8.4cm]{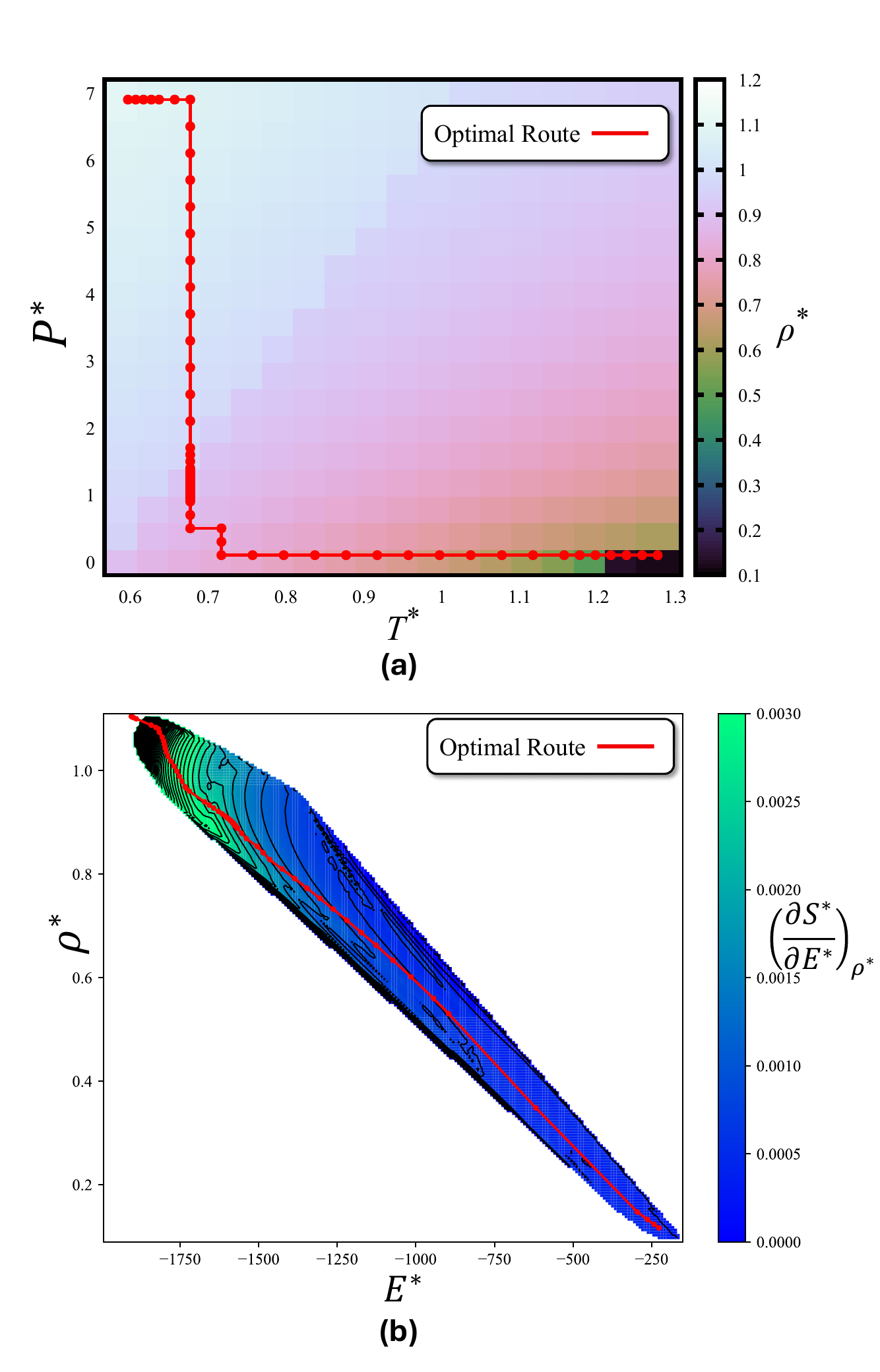}%
\caption{(a) The optimal pathway in temperature and pressure space that traverses the solid-liquid phase boundary, as determined using Dijkstra's algorithm.
(b) The entropy gradient is computed, demonstrating that the optimal pathway shown in (a) follows a smooth gradient.}
\label{result2}
\end{figure}

\begin{figure}[h]
\includegraphics[width=7.4cm]{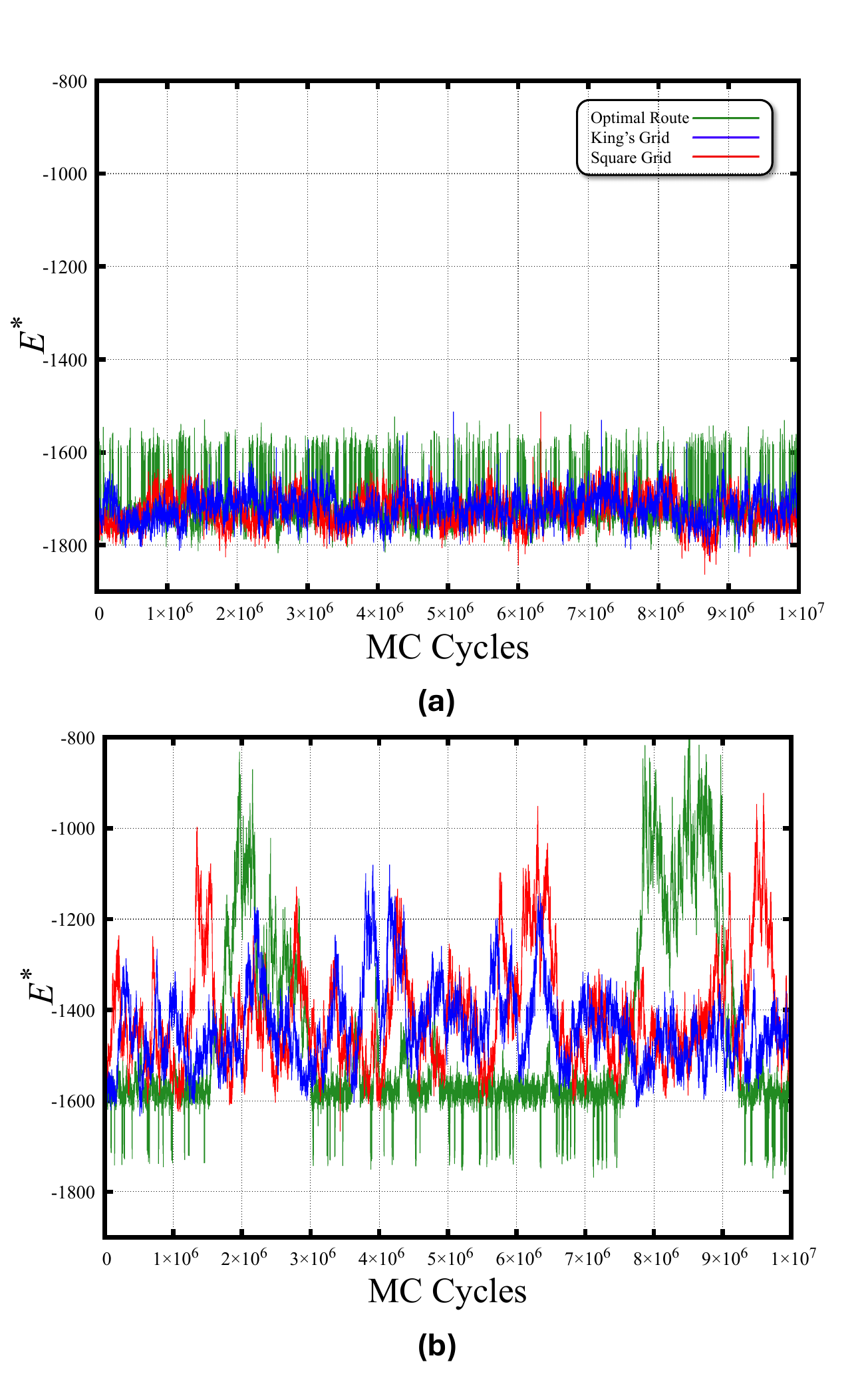}%
\caption{(a) Trajectory of the potential energy $E^*$ versus MC cycles for a replica at whose initial conditions at MC cycle = 0 are $P^* = 1.7$, $T^* = 0.68$ (solid phase).
(b) Trajectory of the potential energy $E^*$ versus MC cycles for a replica at whose initial conditions at MC cycle = 0 are $P^* = 1.3$, $T^* = 0.68$ (liquid phase). In both panels, green, blue, and red lines correspond to simulations using the optimal route, King’s grid, and square grid, respectively.}
\label{result3}
\end{figure}

The results show the identification of the shortest path connecting the solid and gas phases, as shown in Fig. \ref{result2}(a). Since the solid-gas transition represents the most challenging route, spanning from one end of the phase diagram to the other, the obtained path inherently includes the solid-liquid transition. The term ``shortest path'' here refers to the path that minimizes the number of replicas, subject to the constraint that the exchange probability between replicas always exceeds $30\%$. \par
In the replica exchange method, when extracting high-probability pathways from the grid configuration, it is anticipated that the exchange probability will decrease slightly. Therefore, after identifying high-exchange-probability pathways connecting the phases, a short MCMC simulation was performed, and areas with lower exchange probabilities were artificially interpolated to ensure that the exchange probability exceeded $30\%$. (As for the detailed temperature, pressure, and simulation conditions for the optimal path are shown in Supplementary Table S2.) \par
As shown in Fig. \ref{result2}(a), the pathway through temperature-pressure space is nonlinear, and thus the optimal path initially appears to take a detour. However, as indicated by the entropy gradient in eq.\eqref{entropy}, this path is expected to be one where the entropy gradient is relatively smooth. Histograms obtained from the replica exchange Monte Carlo sampling are presented in Fig. \ref{result2}(b), illustrating the trajectory on the entropy gradient plane. The shortest path between the start and ending points obtained aligns with the path where the entropy gradient is minimized as was expected.\par
Next, based on the optimal path calculated using Dijkstra’s algorithm, the trajectories of the potential energy $E^*$ for replicas near the solid-liquid phase transition are shown in Fig. \ref{result3}. Fig. \ref{result3}(a) represents the solid phase replica at $P^* = 1.7$, $T^* = 0.68$, and Fig. \ref{result3}(b) represents the liquid phase replica at $P^* = 1.3$, $T^* = 0.68$. In Fig. \ref{result3}(a), when random walks were performed on the King's grid and Square grid, only the low-energy region within the solid phase was sampled. This observation is consistent with the results shown in Fig. \ref{result1}(a)(c), where these grid configurations sampled only the solid phase. In contrast, the trajectories obtained from the simulation along the optimal path exhibit significant energy fluctuations, indicating that both the solid and liquid phases were successfully sampled.\par
In Fig. \ref{result3}(b), when random walks were conducted on the King's grid and Square grid, the sampling was confined to the liquid energy region. This is also in agreement with the findings in Fig. \ref{result1}(b)(d), where these grid configurations sampled only the liquid phase. However, the trajectories from the simulation along the optimal path demonstrate that the sampling effectively covered the solid phase, liquid phase, and even regions near the gas-liquid phase transition.

\section{Conclusions and Future Work}

In this work, we developed an algorithm that removes the constraints of the conventional replica exchange Monte Carlo method by systematically identifying optimal exchange pathways with the minimal number of replicas. Beyond the standard two-dimensional ``Square grid'' approach, we introduced the ``King's grid,'' which incorporates diagonal exchanges. Through this framework, we investigated the impact of different grid configurations on replica exchange behavior in both solid and liquid phases. The results revealed that each replica tends to remain in either the solid or liquid phase, highlighting the fundamental difficulty of sampling across phase transitions using traditional replica exchange techniques.\par
By leveraging Dijkstra’s algorithm, we successfully identified globally optimal exchange pathways characterized by high-probability transitions for a simple Lennard--Jones bulk system. Additionally, the obtained path indicated that the temperature and pressure intervals along the path correspond to a gradual entropy gradient. This technique offers a novel way to identify the temperature and pressure conditions required to sample a given phase, even when only the state of the phase is known.\par
The approach could also provide deep insights into complex physico-chemical phenomena, such as the simple formation of ice from water, the superconductivity of hydrogen sulphide under certain temperature and pressure conditions, and the separation of mixed solid phases through temperature and pressure pathways that avoid amorphism. These findings suggest that optimizing exchange pathways could serve as a powerful tool for accelerating sampling in systems where conventional methods struggle, ultimately enhancing our ability to explore challenging energy landscapes. Furthermore, the optimization of exchange pathways is not only beneficial for simulations, but may also have direct implications for experimental systems. By identifying efficient phase-transition pathways in the thermodynamic landscape, this approach is expected to help design experimental protocols to control phase transformations and relaxation dynamics through simultaneous control of both the system's temperature and pressure. Such optimization could improve efficiency in industrial processes and have far-reaching benefits.

\section{Acknowledgements}
The authors would like to thank Prof. Shinji Tsuneyuki and Prof. Koji Hukushima for their insightful discussions and valuable feedback.
The computations in this work were carried out using the resources of supercomputer Fugaku provided by the RIKEN Center for Computational Science (Project ID: hp240224). In addition, this work was supported by JSPS KAKENHI Grant Number JP202222673. 

\bibliographystyle{jpc}
\bibliography{sample}

\end{document}


\pagenumbering{arabic}

\title{Supplementary: Navigating Phase Transitions with Game Algorithms: A Strategic Approach to Replica Exchange Monte Carlo}
\author{Akie Kowaguchi}
\affiliation{Department of Mechanical Engineering, Keio University, Yokohama 223-8522, Japan}
\affiliation{Cygames Research, Cygames Inc., Shibuya, Tokyo, Japan}

\author{Katsuhiro Endo}
\affiliation{Research Center for Computational Design of Advanced Functional Materials, National Institute of Advanced Industrial Science and Technology (AIST),
1-1-1 Umezono, Tsukuba, Ibaraki, 305-8568, Japan}

\author{Kentaro Nomura}
\affiliation{Preferred Networks, Inc.
1-6-1 Otemachi, Chiyoda, Tokyo, Japan}

\author{Shuichi Kurabayashi}
\affiliation{Graduate School of Media and Governance, Keio University, Fujisawa, Kanagawa, Japan, }
\affiliation{Cygames Research, Cygames Inc., Shibuya, Tokyo, Japan}
\author{Paul E. Brumby}
\affiliation{Department of Mechanical Engineering, Keio University, Yokohama 223-8522, Japan}
\author{Kenji Yasuoka}
\affiliation{Department of Mechanical Engineering, Keio University, Yokohama 223-8522, Japan}

\maketitle
\setcounter{figure}{0}
 \makeatletter
 \renewcommand{\thefigure}{S\@arabic\c@figure}
 \setcounter{equation}{0}
 \renewcommand{\theequation}{S\@arabic\c@equation}
 \setcounter{table}{0}
 \renewcommand{\thetable}{S\@arabic\c@table}
 \setcounter{section}{0}
 \renewcommand{\thesection}{S\@arabic\c@section}

\section{Dijkstra Algorithm}
The flow of the Dijkstra Algorithm used in this work is shown in Algorithm \ref{Dijkstrasalgorithm}. The Dijkstra method is a process that adds the vertex with the smallest distance in $Q=V-S$ to the set of visited nodes $S$. This method is based on a greedy algorithm, which is the simplest among path finding algorithms, that tries to find a globally optimal solution by repeatedly making the best choices in the current situation. We also show an example of the adjacency matrix shown in Fig.1. 
\begin{algorithm}
\caption{Dijkstra's Algorithm}
\label{Dijkstrasalgorithm}
\begin{algorithmic}[1]
\State \textbf{Initialize:}
\State $S \gets \{\}$ \Comment{Initialize the set of visited nodes.}
\State $Q \gets \{\text{all nodes}\}$ \Comment{Initialize the set of unvisited nodes.}
\For{each node $x$ in $Q$}
    \State $d(x) \gets \infty$
\EndFor
\State $d(s) \gets 0$ \Comment{Initialize the starting point $s$.}
\While{$Q$ is not empty}
    \State Select the nearest node $u$ from $s$
    \State Remove $u$ from $Q$
    \State Add $u$ to $S$
    \For{each neighbor $v$ of $u$ in $Q$}
        \If{$d(v) > d(u) + \text{weight}(u,v)$}
            \State $d(v) \gets d(u) + \text{weight}(u,v)$
        \EndIf
    \EndFor
\EndWhile
\State \textbf{End of Algorithm}
\end{algorithmic}
\end{algorithm}

\begin{figure}[h]
\[
\begin{bmatrix}
1 & 2 & -\log{P_{12}}+\alpha\\
1 & 3 & -\log{P_{13}}+\alpha\\  
2 & 4 & -\log{P_{24}}+\alpha\\  
3 & 4 & -\log{P_{34}}+\alpha\\
\end{bmatrix}
\]

\caption{Definition of the adjacency matrix in the Dijkstra method for the example shown in Fig.1. The adjacency matrix has a dimension of number of nodes * number of nodes and includes all edges that are connected in the vertical and horizontal axes, respectively.}
\label{adjacentmatrix}
\end{figure}

\section{Replica Exchange on Conventional Grids}

The simulation conditions of the replica exchange on conventional grids are given in the table \ref{rootbeforetable}.\par

\begin{table}[ht]
\caption{Simulation Conditions}
\centering
\renewcommand{\arraystretch}{1.1} 
\setlength{\tabcolsep}{8pt} 
\begin{tabular}{@{} p{0.35\columnwidth} p{0.55\columnwidth} @{}} 
\toprule 
\textbf{Exchange Axis} & Temperature $T^*$ and Pressure $P^*$ \\
\midrule 
\textbf{Num. of Replicas} & 18 values of pressure $\times$ 18 values of temperature, total of 324 replicas \\
\midrule
\textbf{Potential} & Lennard--Jones \\
\midrule
\textbf{Num. of Particles} & 256 \\
\midrule
\textbf{Ensemble} & NPT \\
\midrule
\textbf{Phase Transition} & Solid-Liquid and Liquid-Gas \\
\midrule
\textbf{MC Steps} & 20,000,000 \\
\midrule
\textbf{Exchange Frequency} & 1/1000 MC steps \\
\midrule
\textbf{Initial State} & Liquid \\
\midrule
\textbf{Box Length} & 13.63$\sigma$ (gas), 7.17$\sigma$ (liquid), 6.18$\sigma$ (solid) \\
\midrule
\textbf{Cutoff Distance} & 3.00$\sigma$ \\
\bottomrule 
\end{tabular}
\label{rootbeforetable}
\end{table}

To address potential concerns about whether the failure to cross phase boundaries was limited to the specific replicas shown in the main text, we generated heat maps of the replica trajectories across the entire pressure-temperature plane. Using the specific heat peak as a criterion, we labeled the replicas as solid, liquid, or gas in the pressure-temperature space and visualized the trajectories separately for each phase. 
The specific heat capacity is expressed as :
\begin{equation}
\label{heatcapacity}
c_p^{*} = N_{\mathrm{mol}} \frac{<h^{*2}>-<h^{*}>^{2}}{T^{*2}} ,
\end{equation}
where $h^{*}$ is the enthalpy:
\begin{equation}
\label{enthalpy}
h^* = (E^{*} + P^{*}V^*) / N_{\mathrm{mol}}. 
\end{equation}
Here, $E^{*}$ denotes the internal energy, $P^{*}$ is the pressure, $V^{*}$ is the volume, and $N_{\mathrm{mol}}$ is the number of molecules in the system.

\begin{figure*}[h]
\includegraphics[width=18.4cm]{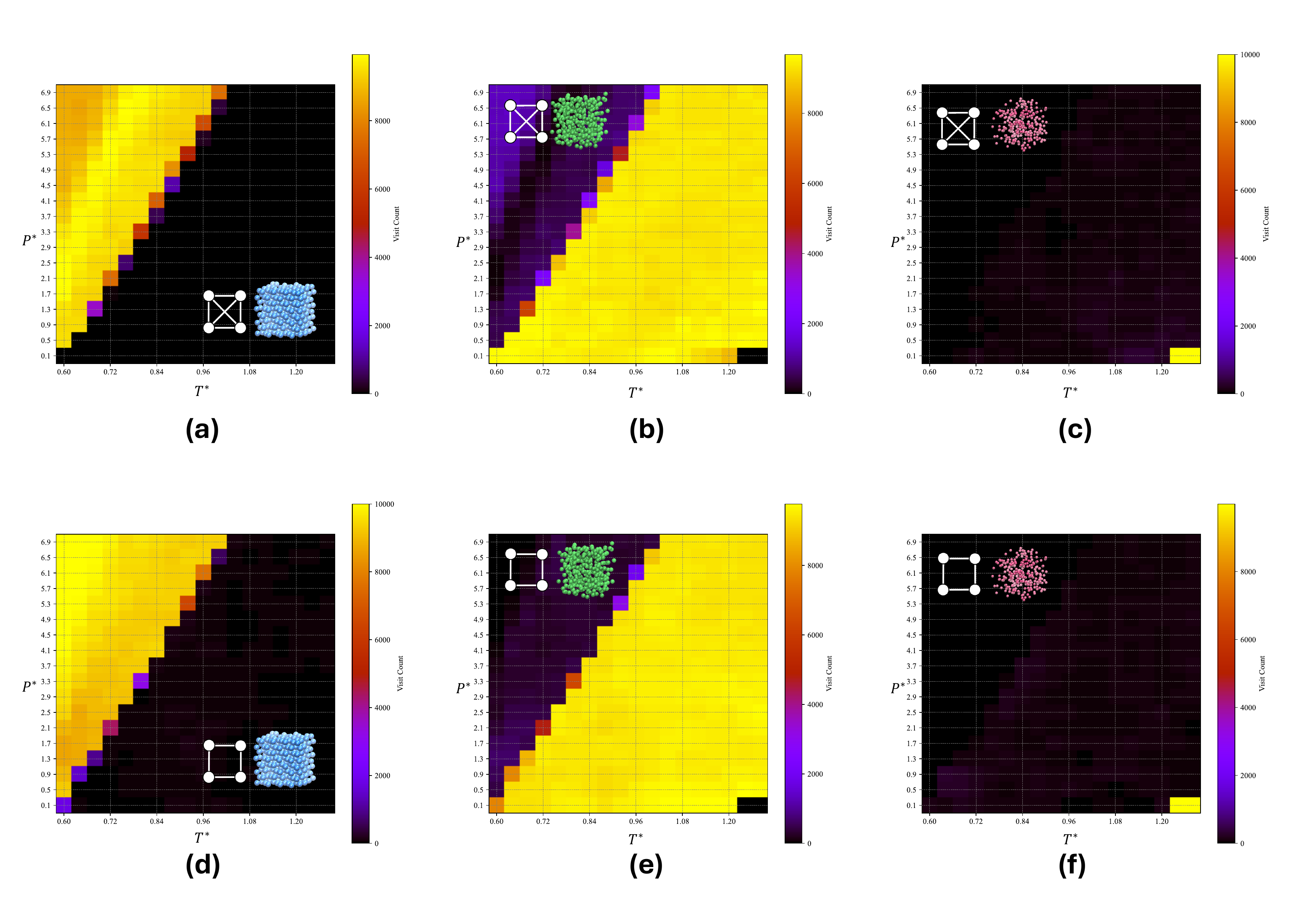}
\caption{Heatmaps of replica trajectories across different initial phases on the King's grid and Square grid. (a) Solid phase on the King's grid, (b) Liquid phase on the King's grid, (c) Gas phase on the King's grid, (d) Solid phase on the Square grid, (e) Liquid phase on the Square grid, and (f) Gas phase on the Square grid.}
\label{heatmap}
\end{figure*}

The results in Fig. \ref{heatmap} confirmed that, regardless of the initial phase, replicas using conventional grid configurations—including both the Square grid (allowing only horizontal and vertical moves) and the King's grid (which incorporates diagonal moves)—failed to cross phase boundaries effectively. This reinforces our conclusion that conventional grids cannot facilitate efficient sampling across phase transitions without pathway optimization.

\section{Optimal route}

The simulation conditions of the optimum route are given intable \ref{rootaftertable}, which are the same as those used for the non-optimal grid conditions, ensuring a fair comparison between the two cases.\par

\begin{table}[ht]
\caption{Simulation Conditions}
\centering
\renewcommand{\arraystretch}{1.1} 
\setlength{\tabcolsep}{8pt} 
\begin{tabular}{@{} p{0.35\columnwidth} p{0.55\columnwidth} @{}} 
\toprule 
\textbf{Exchange Axis} & Temperature $T^*$ and Pressure $P^*$ \\
\midrule 
\textbf{Number of Replicas} & 70 replicas \\
\midrule
\textbf{Potential} & Lennard Jones \\
\midrule
\textbf{Number of Particles} & 256 particles \\
\midrule
\textbf{Ensemble} & NPT \\
\midrule
\textbf{Phase Transition} & Solid-Liquid and Liquid-Gas \\
\midrule
\textbf{MC Steps} & 20,000,000 \\
\midrule
\textbf{Exchange Frequency} & 1/1000 MC steps \\
\midrule
\textbf{Initial State} & Liquid \\
\midrule
\textbf{Box Length} & 13.63$\sigma$ (gas), 7.17$\sigma$ (liquid), 6.18$\sigma$ (solid) \\
\midrule
\textbf{Cutoff Distance} & 3.00$\sigma$ \\
\bottomrule 
\end{tabular}
\label{rootaftertable}
\end{table}

The temperature and pressure conditions of the optimum route and the corresponding replica numbers are given in the following table.\par

\begin{longtable}{|c|c|c|}
\hline
Replica number & Temperature ($T^*$) & Pressure ($P^*$) \\ \hline
1 & 0.6 & 6.9 \\ \hline
2 & 0.61 & 6.9 \\ \hline
3 & 0.62 & 6.9 \\ \hline
4 & 0.63 & 6.9 \\ \hline
5 & 0.64 & 6.9 \\ \hline
6 & 0.66 & 6.9 \\ \hline
7 & 0.68 & 6.9 \\ \hline
8 & 0.68 & 6.5 \\ \hline
9 & 0.68 & 6.1 \\ \hline
10 & 0.68 & 5.7 \\ \hline
11 & 0.68 & 5.3 \\ \hline
12 & 0.68 & 4.9 \\ \hline
13 & 0.68 & 4.5 \\ \hline
14 & 0.68 & 4.1 \\ \hline
15 & 0.68 & 3.7 \\ \hline
16 & 0.68 & 3.3 \\ \hline
17 & 0.68 & 2.9 \\ \hline
18 & 0.68 & 2.5 \\ \hline
19 & 0.68 & 2.1 \\ \hline
20 & 0.68 & 1.7 \\ \hline
21 & 0.68 & 1.6 \\ \hline
22 & 0.68 & 1.5 \\ \hline
23 & 0.68 & 1.4 \\ \hline
24 & 0.68 & 1.38 \\ \hline
25 & 0.68 & 1.36 \\ \hline
26 & 0.68 & 1.34 \\ \hline
27 & 0.68 & 1.32 \\ \hline
28 & 0.68 & 1.3 \\ \hline
29 & 0.68 & 1.28 \\ \hline
30 & 0.68 & 1.26 \\ \hline
31 & 0.68 & 1.24 \\ \hline
32 & 0.68 & 1.22 \\ \hline
33 & 0.68 & 1.2 \\ \hline
34 & 0.68 & 1.18 \\ \hline
35 & 0.68 & 1.16 \\ \hline
36 & 0.68 & 1.14 \\ \hline
37 & 0.68 & 1.12 \\ \hline
38 & 0.68 & 1.1 \\ \hline
39 & 0.68 & 1.08 \\ \hline
40 & 0.68 & 1.06 \\ \hline
41 & 0.68 & 1.04 \\ \hline
42 & 0.68 & 1.02 \\ \hline
43 & 0.68 & 1 \\ \hline
44 & 0.68 & 0.98 \\ \hline
45 & 0.68 & 0.96 \\ \hline
46 & 0.68 & 0.94 \\ \hline
47 & 0.68 & 0.92 \\ \hline
48 & 0.68 & 0.9 \\ \hline
49 & 0.68 & 0.7 \\ \hline
50 & 0.68 & 0.5 \\ \hline
51 & 0.72 & 0.5 \\ \hline
52 & 0.72 & 0.3 \\ \hline
53 & 0.72 & 0.1 \\ \hline
54 & 0.76 & 0.1 \\ \hline
55 & 0.8 & 0.1 \\ \hline
56 & 0.84 & 0.1 \\\hline
57 & 0.88 & 0.1 \\ \hline
58 & 0.92 & 0.1 \\ \hline
59 & 0.96 & 0.1 \\ \hline
60 & 1.00 & 0.1 \\ \hline
61 & 1.04 & 0.1 \\ \hline
62 & 1.08 & 0.1 \\ \hline
63 & 1.12 & 0.1 \\ \hline
64 & 1.16 & 0.1 \\ \hline
65 & 1.18 & 0.1 \\ \hline
66 & 1.20 & 0.1 \\ \hline
67 & 1.22 & 0.1 \\ \hline
68 & 1.24 & 0.1 \\ \hline
69 & 1.26 & 0.1 \\ \hline
70 & 1.28 & 0.1 \\ \hline
\end{longtable}